# Super-Resolution Posterior Ocular Microvascular Imaging Using 3-D Ultrasound Localization Microscopy With a 32×32 Matrix Array

Junhang Zhang, U-Wai Lok , Jingke Zhang, Chengwu Huang, Xin Sun, Chi-Feng Chang, Baoqiang Liu, Chen Gong, Yushun Zeng, Kaipeng Ji, Ryan M. DeRuiter, Jingyi Yin, Lijie Huang, Yanzhe Zhao, Ying Liu, Brian Song, Mark Humanyun, Shigao Chen, Qifa Zhou

*Abstract*—The purpose of this study is to enable *in-vivo* three-dimensional (3-D) ultrasound localization microscopy (ULM) of posterior ocular microvasculature using a 256-channel system and a 1024-element matrix array, and to overcome limitations of restricted transmit angles, sound speed mismatch caused by the crystalline lens and surrounding tissues, and the low signal-to-noise ratio (SNR) of microbubble signals. To address phase distortions from the crystalline lens, which has a higher speed of sound (SOS) than surrounding tissues, a region-dependent SOS beamforming approach was implemented to improve microbubble resolution. A 4-D non-local means filter was subsequently applied to suppress background noise and enhance microbubble contrast. The proposed method improved localization accuracy and image quality, achieving a spatial resolution of 63 μm, while Fourier shell correlation (½-bit threshold) confirmed a global resolution of approximately 59 μm. Higher mean normalized cross-correlation coefficients between the microbubbles and the system point-spread function, obtained with the proposed method (approximately 0.67), compared with those without the proposed method (approximately 0.60), indicate enhanced microbubble signal quality. Furthermore, the 3-D bi-directional vessel density and flow-velocity maps were reconstructed, capturing detailed choroidal vascular and hemodynamic patterns. These results demonstrate that region-dependent SOS beamforming combined with spatiotemporal denoising enables high-resolution posterior ocular ULM and provides a practical pathway toward quantitative 3-D assessment of retinal and choroidal microvasculature for potential clinical use.

## I. INTRODUCTION

THE ability to visualize and functionally assess the posterior segment microvasculature, especially in the retina and choroid, is considered a critical step in early diagnosis, monitoring of disease progression, and assessment of therapeutic responses in the most prevalent ocular conditions affecting diabetic retinopathy[1], [2], [3], glaucoma[4], [5], and age-related macular degeneration[6]. These diseases inherently involve microvascular structural and functional alterations[7], [8]. While common ophthalmic imaging modalities like optical coherence tomography angiography (OCT-A) and fundus fluorescein angiography (FFA) provide high-resolution images of the retinal vascular network, OCT-A still has limited penetration depth[9], [10] that obscures detailed visualization of deeper choroidal vessels and is further impeded by opacities in the ocular media. Although FFA can image deeper vascular structures, it requires the intravenous injection of a fluorescent agent and carries the risk of adverse or allergic reactions[11]. B-mode and Doppler ultrasound are widely used in ophthalmic diagnostics because they provide real-time imaging capability, non-invasive, and have much higher penetration depth compared to optical imaging modalities[12], [13]. However, due to the acoustic diffraction limit, their spatial resolution is limited to several hundred micrometers[14], [15]. This resolution is far from sufficient for precisely visualizing microcirculatory structures with capillary diameters measured by a few micrometers. Consequently, for further understanding of ocular hemodynamics and allowing for precision diagnosis and treatment of ophthalmic diseases, there is an essential need to develop novel imaging modalities that combine sufficient penetration depth with micrometer-scale resolution. To overcome the acoustic diffraction limit and penetration depth,

†Junhang Zhang and U-Wai Lok Equally contributed to this work
U-Wai Lok, Jingke Zhang, Chengwu Huang, Kaipeng Ji, Ryan M. DeRuiter, Jingyi Yin, Lijie Huang, Yanzhe Zhao, and Shigao Chen are with the Department of Radiology, Mayo Clinic, College of Medicine and Science, Rochester, MN, USA (e-mail: Lok.U-Wai@mayo.edu; Zhang.jingke@mayo.edu; Huang.chengwu@mayo.edu; ji.kaipeng@mayo.edu; DeRuiter.Ryan@mayo.edu; yin.jingyi@mayo.edu; huang.lijie@mayo.edu; zhao.yanzhe@mayo.edu; Chen.Shigao@mayo.edu).
Junhang Zhang, Xin Sun, Chi-feng Chang, Chen Gong, and Yushun Zeng are with the Alfred E. Mann Department of Biomedical Engineering, University of Southern California, Los Angeles, CA, USA (e-mail: junhangz@usc.edu; xsun7861@usc.edu; chifengc@usc.edu; chen.gong@usc.edu; yushunze@usc.edu).
Baoqiang Liu, Ying Li, and Brian Song are with USC Roski Eye Institute, Department of Ophthalmology, Keck School of Medicine, University of Southern California, Los Angeles, CA, USA (e-mail: baoqiang@usc.edu; Ying.Liu@med.usc.edu; Brian.Song@med.usc.edu).
Qifa Zhou and Mark Humanyun are with USC Roski Eye Institute, Department of Ophthalmology, Keck School of Medicine, University of Southern California, Los Angeles, CA, USA, and USC Ginsburg Institute for Biomedical Therapeutics, University of Southern California, Los Angeles, CA, USA (e-mail: qifazhou@usc.edu, humanyun@med.usc.edu).
(Corresponding authors: Shigao Chen and Qifa Zhou.)
The study was supported by the National Institutes of Health under award numbers of R01EY35084. The content is solely the responsibility of the authors and does not necessarily represent the official views of the National Institutes of Health. The Mayo Clinic and some of the authors (J.Z., C.H. and S.C.) have a potential financial interest (patents/licensing) related to the technology referenced in the research.

Ultrasound Localization Microscopy (ULM) has emerged that yields micrometer-scale spatial resolution with maintained conventional ultrasound penetration depths. This concept is similar to the optical super-resolution microscopy methods such as photoactivated localization microscopy [16], where super-resolution images are reconstructed by localizing and summing up the centroids of spatially isolated microbubbles (MBs) over thousands of ultrasound frames. With recent advances in ULM [17], [18], [19], its effectiveness has been validated in both preclinical and clinical studies. Despite its great potential, conventional 2D ULM typically requires low concentrations of microbubbles to ensure spatial isolation, resulting in prolonged acquisition times ranging from several to tens of minutes to accumulate enough microbubbles for complete reconstruction of the microvascular. Such prolonged acquisition times reduce the efficiency of ultrasound imaging. To overcome this limitation, MB separation strategies [20] have been proposed to divide high-concentration datasets into sparser subsets. Besides that, a number of alternative methods have been explored to accelerate the acquisition, such as deconvolution[21], null-subtraction [22], inpainting [23], and sparsity-based ultrasound super-resolution hemodynamic imaging (SUSHI)[24]. More recently, deep learning-based ULM methods (Deep-ULM) [25], [26], [27], [28], have emerged using neural networks to extract MB features during training, hence allowing the fast detection of MBs and microvascular reconstruction even for high concentrations of MBs.

Despite the major developments made so far in 2D ULM, vascular networks have complex three-dimensional structures. 2D imaging takes only a cross-sectional projection of the vasculature, hence losing much spatial information. Additionally, it is not capable to compensate the out-of-plane motion, which leads to localization inaccuracies and image distortions. The development of 3D ULM is hence crucial for an intensive and accurate depiction of complex microvascular systems, including those at the posterior eye.

3D ULM offers several distinct advantages over 2D ULM. Most importantly, it allows for an accurate representation of the true complex and tortuous architecture of microvascular networks, which is often misrepresented in 2D projections. Furthermore, by tracking microbubbles through a volume rather than a plane, 3D ULM allows for the measurement of the true 3D velocity vector [29], [30], [31], [32], providing a more accurate quantification of hemodynamic parameters than the projected vectors obtained in 2D ULM. This volumetric approach also improves the robustness of microbubble tracking, as trajectories are less likely to be erroneously terminated by out-of-plane motion which is a significant limitation in 2D ULM.

However, extending ULM from 2D to 3D presents significant technical challenges in ophthalmic implementations. First, using a matrix array probe on a system with limited active channels, such as a 1024-element probe driven by a 256-channel system, reduces the effective aperture and limits the achievable volume acquisition rate. This results in a low signal-to-noise ratio (SNR), rendering weak microbubble signals difficult to detect with high fidelity[33]. Second, the speed of sound within the crystalline lens is considerably higher than that of the surrounding vitreous humor[34]. This mismatch introduces phase aberrations as the ultrasound beam propagates to the posterior segment of the eye, which may degrade the quality and spatial resolution of the resulting microbubbles' signal with limited transmitted angle using a 256-channel system.

To address these challenges, we present a 3D ULM imaging and processing pipeline specifically designed for the posterior microvasculature of the rabbit eye, implemented on a 256-channel ultrasound system with a 32×32 matrix array. Our approach integrates several key innovations to overcome the limitations. We first implement a beamforming with region-dependent speed-of-sound (SOS) to mitigate phase aberrations, thereby improving microbubble signal resolution. We then applied a microbubble separation strategy to reduce acquisition time by allowing for higher MB concentrations. Furthermore, we propose a 4-D non-local means (NLM) filter to suppress background noise. Unlike lower-dimensional NLM implementations [35], the 4-D formulation exploits similarities across both spatial (3-D) and temporal dimensions enabling aggregation of structurally and temporally redundant information.

## II. METHODS AND MATERIALS

### A. Microbubble monitoring and Data acquisition

In this study, a 2D matrix array probe (8 MHz, Vermon S.A., Tours, France) comprising 1024 elements with a pitch of approximately 300 μm was used for data acquisition and real-time monitoring. The probe consists of four independent panels, each composed of 8 × 32 elements. The probe was connected to a Universal Transducer Adapter (UTA) 1024 MUX, which allows for the transmission of four panels with different delays through multiplexing. A Verasonics Vantage 256-channel system (Verasonics Inc., Kirkland, WA, USA) was interfaced with the UTA 1024 MUX to facilitate data acquisition. The transmission strategy adopted a "light" transmission scheme as described in [36], wherein each sub-aperture receives signals not only after the transmission of itself but also after the transmissions of its adjacent apertures. This configuration requires 10 transmissions in total per compound acquisition for each tilted angle. The five steering angles used were: (Azimuth, Elevation) = {(0°, 0°), (−5°, 0°), (5°, 0°), (0°, −5°), (0°, 5°)}. As emphasized in [37], assessing the microbubble (MB) quality before ULM acquisition is critical to optimize imaging settings and timing of data acquisition. Therefore, power Doppler images of the two orthogonal planes, which align with the lateral and elevational axes, were reconstructed and displayed [38] for evaluating the location of MBs and determining the timing of data acquisition, as shown in Fig. 1(a). Following this pre-assessment, radio-frequency (RF) channel data were acquired at a sampling rate of 31.2 MHz, with a pulse repetition frequency (PRF) of 10 kHz and a volumetric imaging rate of 200 Hz, as shown in Fig. 1(b). A total of 7,000 volumes (35 seconds) were collected to reconstruct the ULM volume.

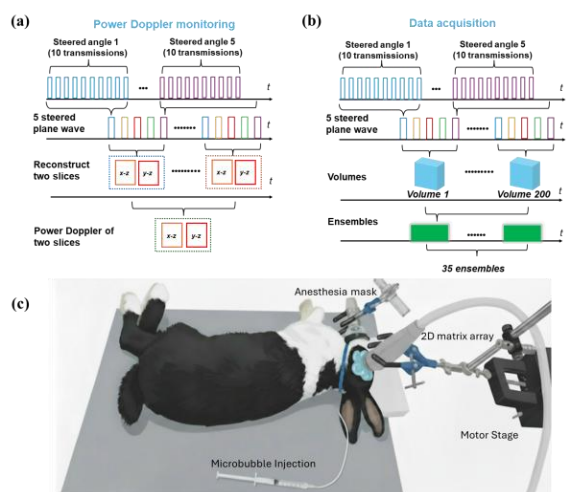

**Fig. 1.** (a) Schematic of power Doppler imaging for two slices for MB evaluation. (b) Schematic of RF channel data acquisition: each volume was compounded using five-angle plane wave transmissions, and 200 volumes of RF data per ensemble were acquired using the Verasonics Vantage 256 system. The sequence was repeated 35 times to collect a total of 7,000 volumes for post-processing. (c) Schematic of the experimental setup for imaging rabbit ocular vasculature.

### B. Ultrasound imaging of the rabbit ocular vasculature

All experimental procedures were conducted on a 5-month-old Dutch Belted rabbit in strict accordance with the guidelines established by the University of Southern California Institutional Animal Care and Use Committee (IACUC) under protocol number 21570. The schematic of the experimental setup is shown in Fig. 1(c). Anesthesia was initially induced via an intramuscular injection of ketamine (35 mg/kg) and xylazine (5 mg/kg) and was subsequently maintained using 2.5% sevoflurane delivered through a facial mask. Throughout the imaging procedure, a heating pad was placed underneath the rabbit to maintain its core body temperature. Prior to imaging, ultrasound gel was applied to the ocular surface to ensure proper acoustic coupling with the transducer. The transducer was mounted on a 3-axis motorized stage (Optosigma, Costa Mesa, CA, USA) for precise positioning and translation, which was used to align the 3D imaging region with the vasculature surrounding the optic nerve head. A 27-gauge needle, attached to a 300μm inner diameter tube, was inserted into the marginal ear vein for microbubble administration. A bolus dose of 100 μL Definity microbubbles (Lantheus, Bedford, MA) diluted with saline at a concentration of 20% was injected, followed by a saline flush immediately after the injection.

### C. Beamforming using Region-Dependent Speed of Sound (RD-SOS)

The crystalline lens exhibits a speed of sound (SOS) of approximately 1730 m/s as reported in [39], which is notably higher than the typical 1540 m/s of soft tissue commonly assumed in conventional ultrasound delay-and-sum beamforming. The mismatch in SOS introduces phase aberrations, leading to resolution degradation in the posterior segment of the eye, particularly under limited transmit angles. The degraded imaging quality is especially detrimental in ULM, where accurate microbubble detection and localization are critical. To address this issue, we implemented a region-dependent speed of sound (RD-SOS) beamforming method. Before applying RD-SOS beamforming to the entire field of view, the SOS of the crystalline lens should be estimated. First, RF channel data were used to reconstruct volumetric data for each angle before compounding using baseband delay-and-sum beamforming with a uniform SOS of 1540 m/s. Next, the upper boundary of the lens was identified, as illustrated in Step 2 of Fig. 2(b). The same dataset was then reconstructed using varying SOS values for both the crystalline lens and the soft tissues below (Step 3) using the RD-SOS beamforming. Finally, the optimal SOS for the lens was determined by identifying the value that maximized the coherence factor across the angular domain data, as shown in Steps 3 and 4. The SOS of the lens was estimated to be 1700 m/s and was subsequently used in the region-dependent RD-SOS beamforming process within the crystalline lens.

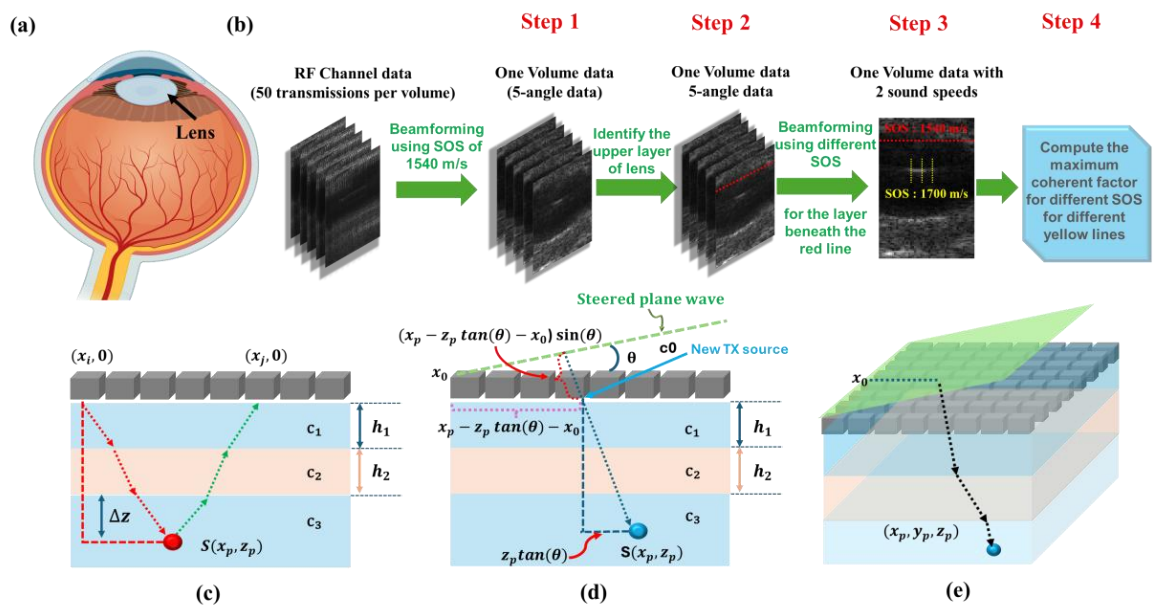

**Fig. 2.** Schematic illustration of the proposed beamforming with region-dependent sound speed estimation framework and regional dependent speed of sound beamforming. (a) Diagram of a rabbit eye, illustrating the lens region, which exhibits a different speed of sound (SOS) compared to surrounding tissues, and posterior ocular microvessel beneath the lens. (b) Step-by-step workflow for estimating the SOS in the lens region. (c) Schematic of the beamforming with different layer model, where ultrasound propagation paths are calculated using different SOS values ($c_1$, $c_2$, $c_3$) in a three-layer structure based on the RD-SOS method. (d) RD-SOS-based beamforming using plane wave imaging. (e) Illustration diagram for 3D plane wave imaging.

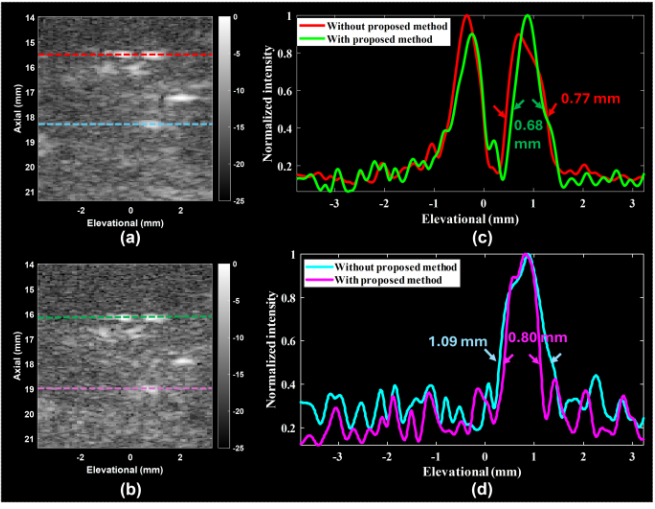

Fig. 3. Elevational maximum-intensity projection (MIP) image (a) without the proposed beamforming method, and (b) with the proposed method, acquired for the centered at a lateral position of 2 mm with a projection thickness of 0.6 mm. (c) Cross-sectional profiles of two MB signals corresponding to the red dashed line in (a) and the green dashed line in (b), used to assess spatial resolution improvement. (d) Cross-sectional profiles of a deeper MB signal along the cyan dashed line in (a) and the magenta dashed line in (b), illustrating that resolution for MB detection can be improved at different depths.

In this study, we assume that the curvature of the lens and the aperture of the 2-D matrix probe are relatively small, allowing the lens region to be approximated as a flat layer. Based on this assumption, our RD-SOS beamforming is a modified version of the multi-layer total focusing method (TFM) that utilizes root-mean-square (RMS) velocity [40]. This method provides rapid and approximate delay calculations based on a hyperbolic time–distance relationship and assumes horizontally layered media, which makes it suitable for GPU processing [41], [42], [43]. Figure 2(c) illustrates a three-layer model in a 2-D configuration, representative of the rabbit eye model (tissue-lens-tissue). The transmit delay from element $i$ to a target scatterer $S$ $(x_p, z_p)$ is depicted by red dashed lines, while the receive delay from the same point to element $j$ is shown in green dashed lines. The time-of-flight along the axial direction from element $i$ to the scatterer $S$ is calculated as:

$$t_0(z_p) = \frac{\Delta z}{c_3} + \frac{h_1}{c_1} + \frac{h_2}{c_2} \quad (1)$$

The root-mean-square velocity can be computed by

$$c_{rms}(z_p) = \sqrt{\frac{h_1 c_1 + h_2 c_2 + c_3 \Delta z}{t_0(z_p)}} \quad (2)$$

Based on the assumptions described in [40] and $z_i = 0$, the root-mean square (RMS) velocity is used to approximate the time-of-flight from transducer element $i$ to the scatterer point $S$, and is given by

$$t(x_p, z_p, x_i, z_i) \approx \sqrt{t_0^2(z_p) + \frac{(x_i - x_p)^2}{c_{rms}^2(z_p)}} \quad (3)$$

The time-of-flight from the scatterer point $S$ to transducer element $j$ (assumed $z_j = 0$) can be calculated in a similar way, and the total time-of-flight since the transmission of element $i$ till the element $j$ receives the backscattered echo from point S is calculated as

$$t(x_p, z_p, x_i, z_i, x_j, z_j) \approx \sqrt{t_0^2(z_p) + \frac{(x_i - x_p)^2}{c_{rms}^2(z_p)}} + \sqrt{t_0^2(z_p) + \frac{(x_j - x_p)^2}{c_{rms}^2(z_p)}} \quad (4)$$

However, this method is only applicable to synthetic aperture imaging, as employed in previous studies. To extend its applicability to plane wave imaging, the transmit time-of-flight must be modified, as illustrated in Fig. 2(d). Specifically, since the transmit source is no longer element $i$, the updated transmit source location must be computed as:

$$x_{i,new} = x_p - z_p \tan\theta - x_0, \quad (5)$$

Secondly, an additional delay must be accounted for due to the steering angle, as indicated by the red-dotted line in Fig. 2(d). The delay corresponding to a given steering angle can be calculated as:

$$t_{extra\ delay} = \frac{(x_p - z_p \tan\theta - x_0)\sin\theta}{c_0} \quad (6)$$

In the 3-D case with a steering angle applied in only one direction, e.g., the lateral direction, as illustrated in Fig. 2(e), the total time-of-flight can be expressed as:

$$t(x_p, y_p, z_p, x_i, y_i, z_i, x_j, y_j, z_j) = \sqrt{t_0^2(z_p) + \frac{(z_p \tan\theta)^2}{c_{rms}^2(z_p)}} + \sqrt{t_0^2(z_p) + \frac{(x_j - x_p)^2 + (y_j - y_p)^2}{c_{rms}^2(z_p)}} + t_{extra\ delay} \quad (7)$$

In this study, the $c_0$, $c_1$, and $c_3$ were set as 1540 m/s, while the $c_2$ was set to 1700 m/s according to the estimated value (see Fig. 2(b)). Additionally, $h_1$ and $h_2$ were set as 3.5 mm and 10.5 mm, respectively.

D. *Motion estimation, tissue clutter filtering and microbubble separation*

For motion estimation, a rigid, sub-pixel motion estimation algorithm was employed. The resulting three-dimensional (3D) displacement values were stored for use in the subsequent MB localization process. Then, a 3D spatiotemporal singular value decomposition-based clutter filter[32] was applied to the post-compounded IQ data to extract moving MB signals from background tissue signals and stationary MB signals. The cutoffs of the tissue clutter subspace were automatically selected using the lower-order singular value thresholding method. Subsequently, an MB separation method was applied to separate original MB data into multiple datasets, each with sparser MB concentration, according to the speed and direction of MB movements. MB signals within the different speeds and motion directions have different Doppler frequencies and can thus be extracted in different subsets.

E. *4D non-local means filtering*

Given a 4-D dataset $I(r)$, where $r = (x,y,z,t)$ indexes spatial and temporal coordinates within a patch. The 4-D non-local means (NLM) filter estimates the denoised MB signals $I(r_0)$ at location $r_0$ as a weighted average of MB signals within a finite search domain $\Omega \in R^4$

$$I(r_0) = \frac{\sum_{r \in \Omega} w(r_0, r) I(r)}{\sum_{r \in \Omega} w(r_0, r)} \quad (8)$$

The similarity weight between the 4-D patch including $r_0$ and $r$ is defined as

$$w(r_0, r) = \frac{1}{Z_{r_0}} \exp\left(-\frac{\|P_{r_0} - P_r\|_2^2}{h^2}\right), \quad (9)$$

where $Z_{r_0}$ denotes the normalization factor ensuring the summation of $w$ to unity, $P_r$ denotes a 4-D patch centered at $r$, $\|.\|_2^2$ is the Euclidean norm, and $h$ is the filtering parameter, with larger h values producing stronger smoothing. Within the search window, the central voxel at $r_0$ is assigned to be the highest weight. The weight magnitude increases with the similarity between the intensity values of the reference patch $P_{r_0}$ and a candidate non-local patch $P_r$. This patch-based comparison enables the algorithm to selectively average only those voxels whose local neighborhoods are structurally similar to that of the reference voxel, thereby preserving fine details while suppressing noise.

*F. 3D MB Localization and tracking*

Each subset of MB signals, after microbubble separation and 4D NLM filtering, was first normalized. The MB signal envelope was then thresholded to suppress background noise. For each volume, motion correction was applied using displacement fields obtained from the preceding motion-estimation step, followed by 3D normalized cross-correlation (NCC) between the MB data and the predefined system-specific point-spread function (PSF), modeled using a multivariate Gaussian function. The NCC map was subsequently thresholded at 0.4 to remove low-correlation pixels, and regional maxima were identified as MB centroids. MB trajectories were then constructed using a 3D bipartition graph–based pairing method with smoothing [32], [44]. This algorithm pairs MB signals by enforcing mutually minimal distances between detections in consecutive frames; two MBs were paired only when this mutual minimum distance criterion was satisfied. A detected MB was considered a reliable signal when it paired across more than five consecutive volumes.

*G. Performance evaluation*

To compare the spatial resolutions with and without RD-SOS beamforming and to evaluate the cross-sectional resolution in the resulting ULM images, vessel cross-sectional profiles were interpolated to a pixel resolution 10 times higher than the original ULM profiles. The full-width at half-maximum (FWHM) of each interpolated cross-sectional vessel profile was then measured to assess the resolution of the super-resolved images. In addition, Fourier shell correlation with the ½-bit threshold was used to obtain a global estimate of the 3D resolution[45].

To compare the contrast enhancement with and without 4D NLM, the contrast ratio (CR) was used to evaluate the improvement with the proposed method in 3D, as

$$CR = 10 \times \log_{10} \frac{MB_{mean}}{N_{mean}}, \quad (10)$$

where the $MB_{mean}$ is the mean intensity including microbubble signals and surrounding signals in the defined region-of-interest (ROI). $N_{mean}$ is the mean background intensity in the defined ROI.

Additionally, successful ULM requires a sufficient number of localized microbubbles (MBs) to enable rapid vessel filling within a short imaging period for *in vivo* applications. The temporal evolution of vessel reconstruction illustrates the progressive development of the vascular network, transitioning from sparse and fragmented structures to a more continuous and saturated vascular map. To quantify this process, a saturation curve was fitted using an exponential saturation model, $C(t) = C_0(1 - e^{-Kt})$, $C(t)$ represents the saturation level at time $t$, $C_0$ is the final saturation level, and $K$ is a rate constant. This model was used to evaluate the efficiency of vessel filling over time.

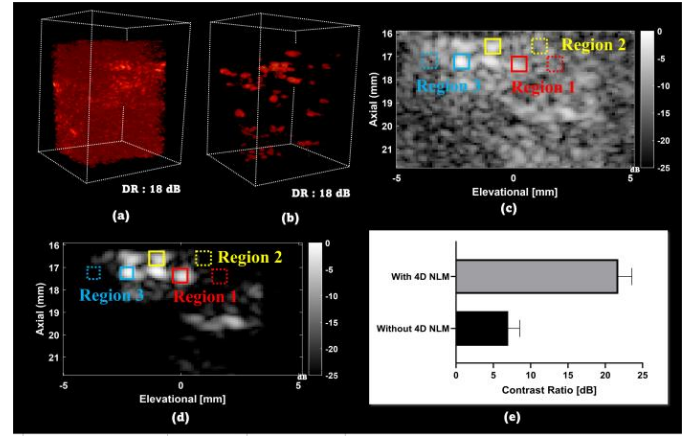

**Fig. 4.** 3D microbubble (MB) images (a) without and (b) with the proposed 4D non-local means (NLM) denoising method. Both 3D images are displayed with an identical dynamic range (DR) of 18 dB for direct comparison. (c) Elevational maximum-intensity projection (MIP) image without 4D NLM filtering at a lateral position of 2 mm and a projection thickness of 0.6 mm. (d) Corresponding MIP image with 4D NLM filtering applied. Solid boxes indicate ROIs used to calculate the mean MB signal intensity, while dashed boxes denote background ROIs for computing the mean background intensity, and (e) CR comparison of the MBs in 20 volumes.

In the MB localization process, MBs were identified as regional maxima of the NCC coefficient map, which reflects the similarity between the MB signals and the reference PSF. Higher NCC maxima indicates better MB signal quality and more accurate localization [37]. Therefore, we evaluate the NCC maximum value of the localizable MBs across 7000 volumes to evaluate MB signal quality with and without the proposed method (RDSOS beamforming and 4D NLM).

### III. RESULTS

*A. Resolution improvement with RD-SOS beamforming*

Figures 3(a) and 3(b) present MB elevational maximum-intensity projection (MIP) images reconstructed without and with the proposed RD-SOS beamforming method, respectively. To quantitatively assess spatial resolution, elevational intensity profiles were extracted along the red dashed line in Fig. 3(a) and the green dashed line in Fig. 3(b), corresponding to the same MB, and are shown in Fig. 3(c). The proposed method yields a profile with a FWHM of 0.68 mm, representing an improvement of 0.09 mm compared to the FWHM of 0.77 mm obtained without the method. Additionally, a deeper MB target is evaluated using cross-sectional profiles indicated by the cyan and magenta dashed lines in Figs. 3(a) and 3(b), with corresponding profiles shown in Fig. 3(d). In this case, the FWHM is reduced from 1.09 mm (without the method) to 0.80 mm (with the proposed method), further demonstrating the proposed RD-SOS beamforming technique's ability to enhance resolution of MB imaging at different depths.

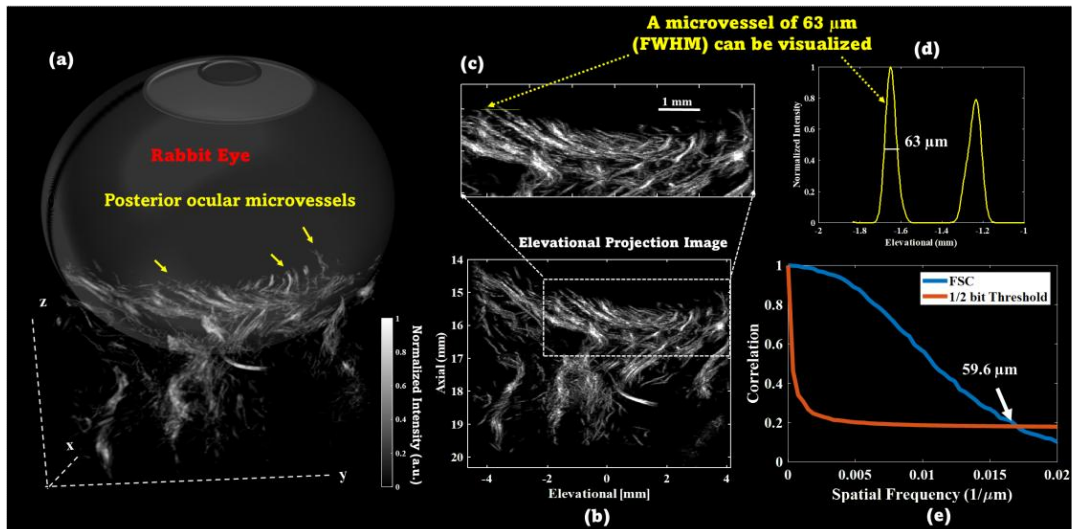

Fig. 5. (a) Three-dimensional ultrasound localization microscopy (ULM) image of the microvasculature. (b) Corresponding elevational maximum-intensity projection (MIP) image of 3D ULM to demonstrate the projection of the vascular structure. (c) Magnified view of the region highlighted by the white dashed box in (b), illustrating the ability of the proposed technique to resolve microvessels with high spatial resolution, (d) The corresponding cross-sectional profiles showing that as small as 63 um blood vessel can be visualized, and (e) Fourier shell correlation (FSC) curve used to estimate the spatial resolution of the reconstructed 3D image. The resolution is determined at the intersection point between the FSC curve and the ½-bit threshold criterion. The eyeball shown in (a) is for illustrative purposes only and does not represent the actual eyeball of the rabbit.

B. *Improved microbubble contrast using 4-D non-local means filtering*

Figures 4(a) and 4(b) present 3D renderings of MB distributions reconstructed from volumetric ultrasound data without and with the proposed 4D NLM denoising method, respectively. Both renderings are visualized with an identical dynamic range (18 dB) to enable direct comparison of noise suppression performance. As shown in Fig. 4(a), the reconstruction without denoising exhibits significant background noise, which hampers the visibility and localization of individual MB signals. In contrast, Fig. 4(b) demonstrates that applying the 4D NLM algorithm substantially suppresses background noise, resulting in clearer MB delineation and improved structural definition.

To further evaluate contrast enhancement, elevational MIP images centered at a lateral position of 2 mm with a projection thickness of 0.6 mm are shown in Figs. 4(c) and 4(d) for the cases without and with 4D NLM filtering, respectively. Three ROIs, marked by red (Region 1), yellow (Region 2), and blue (Region 3) boxes, were selected at varying depths for contrast ratio (CR) analysis. Solid boxes correspond to vessel regions used to compute mean MB signal intensity, while dashed boxes represent adjacent background regions. The corresponding CR values are summarized in Table I, where the proposed 4D NLM method achieved improvements of 12.91 dB, 14.11 dB, and 15.12 dB in regions 1, 2, and 3, respectively.

For statistical validation, CR of 20 MBs was computed across 20 volumes, with results summarized in Fig. 4(e). The mean ± standard deviation of CR increased from $6.98 \pm 1.53$ dB without filtering to $21.75 \pm 1.81$ dB with 4D NLM denoising, an average enhancement of approximately 14 dB. These findings confirm the effectiveness of the proposed 4D NLM denoising technique in improving contrast, thereby enhancing MB detectability and supporting more accurate localization in downstream super-resolution imaging applications.

TABLE I. CRs with and without 4D NLM filtering.

|  | CR (without NLM filtering) | CR (with NLM filtering) |
| --- | --- | --- |
| Region 1 | 5.76 dB | 18.67 dB |
| Region 2 | 6.31 dB | 20.42 dB |
| Region 3 | 9.96 dB | 25.08 dB |

C. *3D super-resolved image of the posterior ocular region and resolution evaluation*

The 3D super-resolution images reconstructed using the proposed method are shown in Fig. 5(a), while the corresponding elevational MIP image is presented in Fig. 5(b). The 3D super-resolution microvessel projection image reveals details that enable visualization of the posterior ocular microvessels (as indicated by the yellow arrow in Fig. 5(a). To further illustrate the resolution enhancement, a small region from the MIP super-resolved image, obtained using the proposed method, was magnified, as shown in Fig. 5(c). Two projected microvessels are indicated by the yellow-dashed arrows. The FWHM of a representative microvessels was measured to be approximately 63 μm as shown in Fig. 5(d). This measurement confirms that the proposed method achieves spatial resolution beyond the diffraction limit, enabling high-resolution delineation of posteriori ocular microvessels.

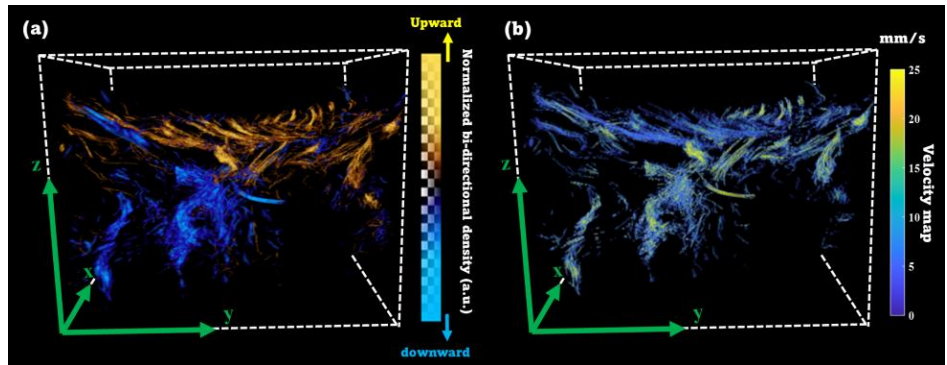

**Fig. 6.** (a) Bi-directional flow map obtained from three-dimensional ultrasound localization microscopy (3D ULM) of the posterior ocular vasculature. Blood vessels with flow toward the ultrasound probe are color-coded in yellow, while those with flow away from the probe are shown in blue, enabling visualization of flow directionality. (b) Corresponding 3D velocity map derived from ULM data, illustrating the spatial distribution of flow velocities and providing further insight into the hemodynamic properties of the ocular microcirculation.

In addition to the local resolution analysis, the Fourier Shell Correlation method was employed to obtain a global estimate of the 3D resolution. For the selected imaging volume, the FSC curve was computed and analyzed using the widely adopted ½-bit threshold criterion. As shown in Fig. 5(e), the spatial resolution corresponding to the intersection point of the FSC curve and the ½-bit threshold was determined to be approximately 59 μm.

### D. Bi-directional and velocity evaluation of 3-D super-resolved image of the rabbit eye vasculature

To differentiate flow direction within the posterior ocular vascular network, a 3D bi-directional super-resolved imaging was employed, as illustrated in Fig. 6 (a). In this visualization, blood vessels with flow directed toward the ultrasound probe are color-coded in yellow, while vessels with flow directed away from the probe are shown in blue. The resulting bi-directional image reveals that the retinal/choroidal microvasculature primarily exhibits upward flow, corresponding to arterial perfusion directed toward the inner retinal layers. In contrast, the larger vessels in the subretinal region exhibit downward flow, likely associated with venous drainage or choroidal circulation. The 3D bi-directional super-resolved imaging thus provides a comprehensive depiction of both the vascular morphology and the directional flow characteristics within the posterior segment of the eye.

In addition, 3D velocity-resolved microvascular imaging was performed, as shown in Fig. 6 (b), to further characterize the hemodynamic properties of the ocular circulation. These images visualize the distribution and perfusion of microbubbles within the vascular network, enabling concurrent assessment of both structural and functional vascular information. The resulting velocity maps indicate that retinal/choroidal microvessels predominantly exhibit lower flow velocities, consistent with the fine capillary network at this region. In contrast, the larger subretinal vessels demonstrate significantly higher flow speeds, reflecting their role in rapid blood transport. Together, these advanced imaging modalities offer detailed insights into the structural and functional organization of the ocular microcirculation, which may be critical for early diagnosis and monitoring of retinal and choroidal vascular disorders.

The velocity distribution was evaluated using the velocity map, which delineates distinct flow patterns in the posterior ocular vasculature between the retinal/choroidal layer and the subretinal region. Velocity histograms were computed from two regions of interest (ROIs) indicated in Fig. 7(a): the magenta dashed box (depth: 14–16.5 mm) corresponding to the retinal layer, and the gray dashed box (depth: 16.5–20.5 mm) corresponding to the region beneath the retina. For the retinal/choroidal ROI, the mean ± standard deviation velocities were 9.49 ± 5.86 mm/s. For the region beneath the retina/choroidal, the mean ± standard deviation velocities were 14.25 ± 8.04 mm/s. These measurements are consistent with the presence of smaller, slower-flow microvessels in the retinal/choroidal layer and larger, higher-flow vessels in the subretinal region, as shown in the histogram in Fig. 7(b).

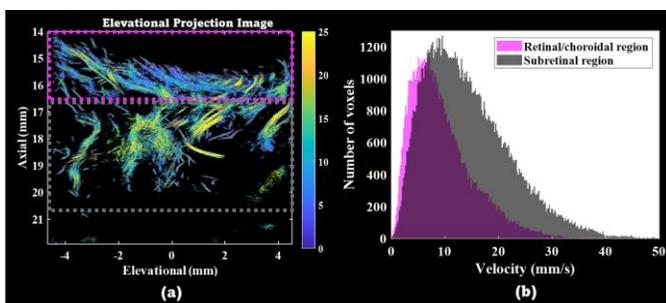

**Fig. 7.** Velocity-resolved analysis of the posterior ocular microvasculature. (a) Elevational maximum-intensity projection (MIP) image color-coded by flow velocity, obtained from 3D ultrasound localization microscopy (ULM) data. The velocity map enables simultaneous visualization of vascular morphology and flow dynamics. Dashed magenta and gray boxes delineate regions corresponding to the retina/choroidal and the subretinal, respectively. (b) Histogram of flow velocities for the retinal region (magenta) and the region in the subretinal (gray), showing a higher prevalence of slower flows in the retina/choroidal microvessels and faster flows in the subretinal region.

### E. Vessel reconstruction over time

To assess the adequacy of this acquisition time, different numbers of volumes were used to reconstruct the 3D ULM, ranging from 1000 to 5000 volumes, as illustrated in MIP images, as shown in Figs. 8(a)–(c). The number of reconstructed microvascular voxels was modeled using an exponential saturation function as shown in Fig. 8(d). Under the current settings, approximately 85% of the vascular network was recovered with 7000 volumes (35 seconds).

### F. Normalized cross correlation with and without the proposed method

Elevational MIP images obtained from 3D ULM images near the retinal/choroidal region without and with the proposed method (RDSOS beamforming + 4D NLM) are shown in Figs. 9(a) and 9(b), respectively. Some small vessels that cannot be visualized without the proposed method become visible with the proposed method, as indicated by the red arrows in Fig 9(b). Additionally, Fig. 9(c) presents the corresponding NCC measurements. Without the proposed method, the mean ± standard deviation of the NCC was 0.602 ± 0.021, whereas with the proposed method it increased to 0.673 ± 0.022. The higher NCC peak achieved using the proposed method indicates improved MB signal quality, leading to more accurate localization.

## IV. Discussion

In this study, we demonstrated the feasibility of visualizing the posterior ocular microvasculature implemented on a 256-channel ultrasound system with a 32×32 matrix array probe. The RD-SOS beamforming method was applied to compensate for the phase aberration caused by the mismatched SOS in the crystalline-lens media, resulting in improved spatial resolution of MB imaging. In addition, a 4-D non-local means filtering was employed to suppress background noises, thereby further enhancing microbubble detectability. With these processing steps, a single projected microvessel with a FWHM of 63 μm was resolved, significantly surpassing the theoretical diffraction-limited resolution of approximately 200 μm (one wavelength). A global resolution estimate obtained via Fourier shell correlation analysis yielded a value of 59.6 μm using the ½-bit threshold criterion. Furthermore, the increased NCC peak achieved with the proposed method demonstrates enhanced MB signal quality, which contributes to more precise localization. Additionally, the proposed 3-D ULM also enables quantitative assessment of flow direction and velocity, thereby facilitating detailed hemodynamic evaluation of the posterior ocular region and enabling detection of changes in blood circulation that may be indicative of disease progression. These capabilities highlight the potential of 3-D ULM as a powerful noninvasive imaging tool for monitoring ocular vascular diseases such as glaucoma, diabetic retinopathy, and age-related macular degeneration.

This study has several limitations that should be acknowledged. First, the limited acoustic signal power inherent to the ultrasound system reduces the SNR for microbubble detection, particularly at greater imaging depths. This limitation degrades microbubble signal quality at deeper regions. To mitigate this issue, a 4-D NLM filtering approach was employed to enhance contrast resolution and suppress noise while preserving microbubbles' signals. Although this post-processing step improved the overall quality of the microbubble data, further performance gains could be achieved through transmission designs with higher penetration efficiency or advanced noise suppression techniques. For example, coded excitation strategies, such as Hadamard coding [46], could be utilized to improve penetration depth and SNR without compromising spatial resolution for 3D imaging.

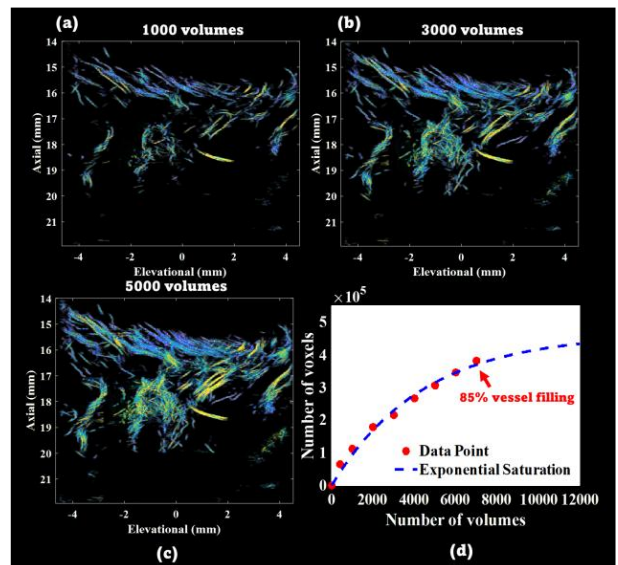

**Fig. 8.** Vessel filling analysis in 3D ULM. Maximum-intensity projection (MIP) images reconstructed using different numbers of volumes: (a) 1000 (5 seconds), (b) 3000 (15 seconds), (c) 5000 (25 seconds). The vascular network gradually evolves from sparse structures to a more continuous and saturated representation as the acquisition time increases. (d) Exponential saturation fitting of the number of localized vascular voxels as a function of the number of volumes. Approximately 85% vessel filling is achieved with 7000 volumes, while around 10,000 volumes (corresponding to at least 50 seconds) are required to reach 95% filling.

Secondly, a fixed sound speed with a flat layer was assumed within the region containing the crystalline lens. While this simplification facilitates image reconstruction and improves the quality of microbubble signals, it does not account for residual phase aberrations in the posterior ocular region. To address these aberrations, recent methods based on deep learning approaches [47] or differentiable beamforming techniques [48] can be employed. However, the spatial invariance assumption inherent in the phase screen model may introduce significant physical inaccuracies in the presence of strongly aberrating layers. Alternatively, patch-based phase aberration correction methods, which divide the imaging domain into multiple patches based on the iso-planatic limit, can be used. In this approach, aberrations are estimated within each patch using metrics such as the coherence factor [49] or mid-angle phase differences [50]. Furthermore, beamforming at a single pixel can be extended to a local reflection point spread function (RPSF), which estimates phase aberration at the pixel level using a distortion matrix [51] and [52]. The proposed beamforming strategy, combined with phase aberration correction, has the potential to further enhance spatial resolution and enable higher-fidelity MB quality for the ocular microvasculature imaging. Furthermore, higher microbubble (MB) concentrations compromise localization accuracy due to overlapping MB signals. In contrast, lower MB concentrations enhance localization precision but require longer acquisition times to achieve sufficient vessel coverage. To shorten the data acquisition time, MB separation approach was used and the vessel filling was evaluated in this study. With the proposed method, approximately 85% vessel coverage can be achieved within 35 seconds of acquisition. While this acquisition time is acceptable in a research setting, it may exceed the duration of a single breath-hold or eye blink, potentially limiting its practicality for clinical translation.

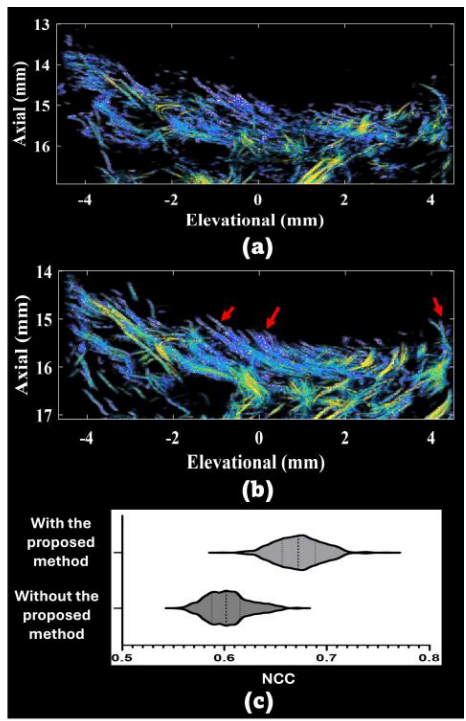

**Fig. 9.** Elevational maximum-intensity projection images (near the retinal/choroidal region) color-coded by flow velocity, obtained from 3D ultrasound localization microscopy (ULM) data: (a) without the proposed RDSOS beamforming and 4D NLM, (b) with the proposed RDSOS beamforming and 4D NLM, and (c) normalized cross-correlation results without and with the proposed method.

Finally, the matrix array used in this study had an aperture size of 9.3 × 10.1 mm, which was insufficient to encompass the entire eyeball and posterior ocular tissue. Expanding the number of array elements could provide a wider field of view; however, such an approach poses significant challenges for transducer fabrication and increases the complexity of the associated electronic system. As an alternative, diverging wave or row–column [53] addressing schemes combined with or without a diverging lens may offer an effective solution, as they can extend the achievable field of view[54].

## V. Conclusion

This study demonstrates that region-dependent SOS beamforming, combined with spatiotemporal non-local means denoising and our proposed ULM post-processing, enables high-resolution 3-D ULM of posterior ocular microvasculature using a 256-channel matrix array system. By mitigating phase distortions from the crystalline lens and enhancing microbubble signal quality, the proposed framework improves localization accuracy and overall image quality. The resulting 3-D vascular and flow-velocity reconstructions capture detailed retinal/choroidal microvascular structures and hemodynamics. These findings highlight the feasibility of quantitative 3-D assessment of posterior ocular microvasculature, with promising implications for future clinical evaluation and monitoring of ocular diseases.

## References

[1] Y. Wu *et al.*, "Early detection of retinal and choroidal microvascular impairments in diabetic patients with myopia," *Front. Cell Dev. Biol.*, vol. 13, p. 1609928, May 2025, doi: 10.3389/fcell.2025.1609928.

[2] C. Yin, S. Zhang, D. Guo, J. Qin, H. Lou, and G. Zhang, "Diagnostic value of choroidal vascular density in predicting the progression of diabetic retinopathy," *Sci Rep*, vol. 15, no. 1, p. 15671, May 2025, doi: 10.1038/s41598-025-00528-y.

[3] H. Nouri, S.-H. Abtahi, M. Mazloumi, S. Samadikhadem, J. F. Arevalo, and H. Ahmadieh, "Optical coherence tomography angiography in diabetic retinopathy: A major review," *Survey of Ophthalmology*, vol. 69, no. 4, pp. 558–574, July 2024, doi: 10.1016/j.survophthal.2024.03.004.

[4] R. Nascimento E Silva *et al.*, "Microvasculature of the Optic Nerve Head and Peripapillary Region in Patients With Primary Open-Angle GlaucomaMicrovasculature of the Optic Nerve Head and Peripapillary Region in Patients With Primary Open-Angle Glaucoma," *Journal of Glaucoma*, vol. 28, no. 4, pp. 281–288, Apr. 2019, doi: 10.1097/IJG.0000000000001165.

[5] X. Fan *et al.*, "The characteristics of fundus microvascular alterations in the course of glaucoma: a narrative review," *Ann Transl Med*, vol. 10, no. 9, pp. 527–527, May 2022, doi: 10.21037/atm-21-5695.

[6] S. Fragiotta *et al.*, "Choroidal Vasculature Changes in Age-Related Macular Degeneration: From a Molecular to a Clinical Perspective," *IJMS*, vol. 23, no. 19, p. 12010, Oct. 2022, doi: 10.3390/ijms231912010.

[7] X. Wei, P. K. Balne, K. E. Meissner, V. A. Barathi, L. Schmetterer, and R. Agrawal, "Assessment of flow dynamics in retinal and choroidal microcirculation," *Survey of Ophthalmology*, vol. 63, no. 5, pp. 646–664, Sept. 2018, doi: 10.1016/j.survophthal.2018.03.003.

[8] P. A. Campochiaro, "Retinal and Choroidal Vascular Diseases: Past, Present, and Future: The 2021 Proctor Lecture," *Invest. Ophthalmol. Vis. Sci.*, vol. 62, no. 14, p. 26, Nov. 2021, doi: 10.1167/iovs.62.14.26.

[9] A. H. Kashani *et al.*, "Optical coherence tomography angiography: A comprehensive review of current methods and clinical applications," *Progress in Retinal and Eye Research*, vol. 60, pp. 66–100, Sept. 2017, doi: 10.1016/j.preteyeres.2017.07.002.

[10] J. Wang, T. T. Hormel, S. T. Bailey, T. S. Hwang, D. Huang, and Y. Jia, "Signal attenuation-compensated projection-resolved OCT angiography," *Biomed. Opt. Express*, vol. 14, no. 5, p. 2040, May 2023, doi: 10.1364/BOE.483835.

[11] I. S. Kornblau and J. F. El-Annan, "Adverse reactions to fluorescein angiography: A comprehensive review of the literature," *Survey of Ophthalmology*, vol. 64, no. 5, pp. 679–693, Sept. 2019, doi: 10.1016/j.survophthal.2019.02.004.

[12] K. S. Brunell, "Ophthalmic Ultrasonography," *Journal of Diagnostic Medical Sonography*, vol. 30, no. 3, pp. 136–142, May 2014, doi: 10.1177/8756479314531333.

[13] R. H. Silverman, "Principles of ophthalmic ultrasound," *Expert Review of Ophthalmology*, vol. 18, no. 6, pp. 379–389, Nov. 2023, doi: 10.1080/17469899.2023.2277781.

[14] K. Christensen-Jeffries *et al.*, "Super-resolution Ultrasound Imaging," *Ultrasound in Medicine & Biology*, vol. 46, no. 4, pp. 865–891, Apr. 2020, doi: 10.1016/j.ultrasmedbio.2019.11.013.

[15] S. Dencks and G. Schmitz, "Ultrasound localization microscopy," *Zeitschrift für Medizinische Physik*, vol. 33, no. 3, pp. 292–308, Aug. 2023, doi: 10.1016/j.zemedi.2023.02.004.

[16] E. Betzig *et al.*, "Imaging Intracellular Fluorescent Proteins at Nanometer Resolution," *Science*, vol. 313, no. 5793, pp. 1642–1645, Sept. 2006, doi: 10.1126/science.1127344.

[17] C. Errico *et al.*, "Ultrafast ultrasound localization microscopy for deep super-resolution vascular imaging," *Nature*, vol. 527, no. 7579, pp. 499–502, Nov. 2015, doi: 10.1038/nature16066.

[18] S. Schwarz *et al.*, "Ultrasound Super-Resolution Imaging of Neonatal Cerebral Vascular Reorganization," *Advanced Science*, vol. 12, no. 12, p. 2415235, Mar. 2025, doi: 10.1002/advs.202415235.

[19] W. Zhang, M. R. Lowerison, Z. Dong, R. J. Miller, K. A. Keller, and P. Song, "Super-Resolution Ultrasound Localization Microscopy on a Rabbit Liver VX2 Tumor Model: An Initial Feasibility Study,"

*Ultrasound in Medicine & Biology*, vol. 47, no. 8, pp. 2416–2429, Aug. 2021, doi: 10.1016/j.ultrasmedbio.2021.04.012.

[20] C. Huang et al., "Short Acquisition Time Super-Resolution Ultrasound Microvessel Imaging via Microbubble Separation," *Sci Rep*, vol. 10, no. 1, p. 6007, Apr. 2020, doi: 10.1038/s41598-020-62898-9.

[21] U.-W. Lok et al., "Improved Ultrasound Microvessel Imaging Using Deconvolution with Total Variation Regularization," *Ultrasound in Medicine & Biology*, vol. 47, no. 4, pp. 1089–1098, Apr. 2021, doi: 10.1016/j.ultrasmedbio.2020.12.025.

[22] Z. Kou, M. R. Lowerison, Q. You, Y. Wang, P. Song, and M. L. Oelze, "High-Resolution Power Doppler Using Null Subtraction Imaging," *IEEE Trans. Med. Imaging*, vol. 43, no. 9, pp. 3060–3071, Sept. 2024, doi: 10.1109/TMI.2024.3383768.

[23] Q. You et al., "Curvelet Transform-Based Sparsity Promoting Algorithm for Fast Ultrasound Localization Microscopy," *IEEE Trans. Med. Imaging*, vol. 41, no. 9, pp. 2385–2398, Sept. 2022, doi: 10.1109/TMI.2022.3162839.

[24] A. Bar-Zion, O. Solomon, C. Tremblay-Darveau, D. Adam, and Y. C. Eldar, "SUSHI: Sparsity-Based Ultrasound Super-Resolution Hemodynamic Imaging," *IEEE Trans. Ultrason., Ferroelect., Freq. Contr.*, vol. 65, no. 12, pp. 2365–2380, Dec. 2018, doi: 10.1109/TUFFC.2018.2873380.

[25] U.-W. Lok et al., "Fast super-resolution ultrasound microvessel imaging using spatiotemporal data with deep fully convolutional neural network," *Phys. Med. Biol.*, vol. 66, no. 7, p. 075005, Apr. 2021, doi: 10.1088/1361-6560/abeb31.

[26] R. J. G. Van Sloun et al., "Super-Resolution Ultrasound Localization Microscopy Through Deep Learning," *IEEE Trans. Med. Imaging*, vol. 40, no. 3, pp. 829–839, Mar. 2021, doi: 10.1109/TMI.2020.3037790.

[27] X. Chen, M. R. Lowerison, Z. Dong, A. Han, and P. Song, "Deep Learning-Based Microbubble Localization for Ultrasound Localization Microscopy," *IEEE Trans. Ultrason., Ferroelect., Freq. Contr.*, vol. 69, no. 4, pp. 1312–1325, Apr. 2022, doi: 10.1109/TUFFC.2022.3152225.

[28] B. Rauby, P. Xing, M. Gasse, and J. Provost, "Deep Learning in Ultrasound Localization Microscopy: Applications and Perspectives," *IEEE Trans. Ultrason., Ferroelect., Freq. Contr.*, vol. 71, no. 12: Breaking the Resolution, pp. 1765–1784, Dec. 2024, doi: 10.1109/TUFFC.2024.3462299.

[29] B. Heiles et al., "Ultrafast 3D Ultrasound Localization Microscopy Using a 32 X 32 Matrix Array," *IEEE Trans. Med. Imaging*, vol. 38, no. 9, pp. 2005–2015, Sept. 2019, doi: 10.1109/TMI.2018.2890358.

[30] G. Chabouh et al., "3D transcranial ultrasound localization microscopy in awake mice: protocol and open-source pipeline," *Commun Eng*, vol. 4, no. 1, p. 102, June 2025, doi: 10.1038/s44172-025-00415-4.

[31] P. Xing et al., "3D ultrasound localization microscopy of the nonhuman primate brain," *eBioMedicine*, vol. 111, p. 105457, Jan. 2025, doi: 10.1016/j.ebiom.2024.105457.

[32] U.-W. Lok et al., "Three-Dimensional Ultrasound Localization Microscopy with Bipartite Graph-Based Microbubble Pairing and Kalman-Filtering-Based Tracking on a 256-Channel Verasonics Ultrasound System with a 32 × 32 Matrix Array," *J. Med. Biol. Eng.*, vol. 42, no. 6, pp. 767–779, Dec. 2022, doi: 10.1007/s40846-022-00755-y.

[33] S. Harput et al., "3-D Super-Resolution Ultrasound Imaging With a 2-D Sparse Array," *IEEE Trans. Ultrason., Ferroelect., Freq. Contr.*, vol. 67, no. 2, pp. 269–277, Feb. 2020, doi: 10.1109/TUFFC.2019.2943646.

[34] T. Matéo, Y. Mofid, J.-M. Grégoire, and F. Ossant, "An Eye-adapted Beamforming for Axial B-scans Free from Crystalline Lens Aberration: In vitro and ex vivo Results with a 20MHz Linear Array," *Physics Procedia*, vol. 70, pp. 1250–1254, 2015, doi: 10.1016/j.phpro.2015.08.278.

[35] P. Song et al., "Improved Super-Resolution Ultrasound Microvessel Imaging With Spatiotemporal Nonlocal Means Filtering and Bipartite Graph-Based Microbubble Tracking," *IEEE Trans. Ultrason., Ferroelect., Freq. Contr.*, vol. 65, no. 2, pp. 149–167, Feb. 2018, doi: 10.1109/TUFFC.2017.2778941.

[36] A. Chavignon, B. Heiles, V. Hingot, C. Orset, D. Vivien, and O. Couture, "3D Transcranial Ultrasound Localization Microscopy in the Rat Brain With a Multiplexed Matrix Probe," *IEEE Trans. Biomed. Eng.*, vol. 69, no. 7, pp. 2132–2142, July 2022, doi: 10.1109/TBME.2021.3137265.

[37] C. Huang et al., "Optimizing *in vivo* data acquisition for robust clinical microvascular imaging using ultrasound localization microscopy," *Phys. Med. Biol.*, vol. 70, no. 7, p. 075017, Apr. 2025, doi: 10.1088/1361-6560/adc0de.

[38] U.-W. Lok et al., "Real time SVD-based clutter filtering using randomized singular value decomposition and spatial downsampling for micro-vessel imaging on a Verasonics ultrasound system," *Ultrasonics*, vol. 107, p. 106163, Sept. 2020, doi: 10.1016/j.ultras.2020.106163.

[39] C. Görig, T. Varghese, T. Stiles, J. Van Den Broek, J. A. Zagzebski, and C. J. Murphy, "Evaluation of acoustic wave propagation velocities in the ocular lens and vitreous tissues of pigs, dogs, and rabbits," *American Journal of Veterinary Research*, vol. 67, no. 2, pp. 288–295, Feb. 2006, doi: 10.2460/ajvr.67.2.288.

[40] X. Guo and Y. Han, "Ultrasonic Total Focusing Imaging Method of Multilayer Composite Structures Using the Root-Mean-Square (RMS) Velocity," *Advances in Materials Science and Engineering*, vol. 2021, no. 1, p. 2745732, Jan. 2021, doi: 10.1155/2021/2745732.

[41] D. Hyun, Y. L. Li, I. Steinberg, M. Jakovljevic, T. Klap, and J. J. Dahl, "An Open Source GPU-Based Beamformer for Real-Time Ultrasound Imaging and Applications," in *2019 IEEE International Ultrasonics Symposium (IUS)*, Glasgow, United Kingdom: IEEE, Oct. 2019, pp. 20–23. doi: 10.1109/ULTSYM.2019.8926193.

[42] U.-W. Lok and P.-C. Li, "Transform-Based Channel-Data Compression to Improve the Performance of a Real-Time GPU-Based Software Beamformer," *IEEE Trans. Ultrason., Ferroelect., Freq. Contr.*, vol. 63, no. 3, pp. 369–380, Mar. 2016, doi: 10.1109/TUFFC.2016.2519441.

[43] B. Y. S. Yiu, I. K. H. Tsang, and A. C. H. Yu, "GPU-based beamformer: Fast realization of plane wave compounding and synthetic aperture imaging," *IEEE Trans. Ultrason., Ferroelect., Freq. Contr.*, vol. 58, no. 8, pp. 1698–1705, Aug. 2011, doi: 10.1109/TUFFC.2011.1999.

[44] P. Song et al., "Improved Super-Resolution Ultrasound Microvessel Imaging With Spatiotemporal Nonlocal Means Filtering and Bipartite Graph-Based Microbubble Tracking," *IEEE Trans. Ultrason., Ferroelect., Freq. Contr.*, vol. 65, no. 2, pp. 149–167, Feb. 2018, doi: 10.1109/TUFFC.2017.2778941.

[45] V. Hingot, A. Chavignon, B. Heiles, and O. Couture, "Measuring Image Resolution in Ultrasound Localization Microscopy," *IEEE Trans. Med. Imaging*, vol. 40, no. 12, pp. 3812–3819, Dec. 2021, doi: 10.1109/TMI.2021.3097150.

[46] J. Zhang et al., "Enhancing Row-Column Array (RCA)-Based 3D Ultrasound Vascular Imaging With Spatial-Temporal Similarity Weighting," *IEEE Trans. Med. Imaging*, vol. 44, no. 1, pp. 297–309, Jan. 2025, doi: 10.1109/TMI.2024.3439615.

[47] M. Sharifzadeh, S. Goudarzi, A. Tang, H. Benali, and H. Rivaz, "Mitigating Aberration-Induced Noise: A Deep Learning-Based Aberration-to- Aberration Approach," *IEEE Trans. Med. Imaging*, vol. 43, no. 12, pp. 4380–4392, Dec. 2024, doi: 10.1109/TMI.2024.3422027.

[48] B. Hériard-Dubreuil, A. Besson, C. Cohen-Bacrie, and J.-P. Thiran, "A Path-Based Model for Aberration Correction in Ultrasound Imaging," *IEEE Trans. Med. Imaging*, vol. 44, no. 8, pp. 3222–3232, Aug. 2025, doi: 10.1109/TMI.2025.3562011.

[49] G. Montaldo, M. Tanter, and M. Fink, "Time Reversal of Speckle Noise," *Phys. Rev. Lett.*, vol. 106, no. 5, p. 054301, Feb. 2011, doi: 10.1103/PhysRevLett.106.054301.

[50] W. Simson, L. Zhuang, B. N. Frey, S. J. Sanabria, J. J. Dahl, and D. Hyun, "Ultrasound Autofocusing: Common Midpoint Phase Error Optimization via Differentiable Beamforming," *IEEE Trans. Med. Imaging*, pp. 1–1, 2025, doi: 10.1109/TMI.2025.3607875.

[51] W. Lambert, L. A. Cobus, T. Frappart, M. Fink, and A. Aubry, "Distortion matrix approach for ultrasound imaging of random scattering media," *Proc. Natl. Acad. Sci. U.S.A.*, vol. 117, no. 26, pp. 14645–14656, June 2020, doi: 10.1073/pnas.1921533117.

[52] F. Bureau, J. Robin, A. Le Ber, W. Lambert, M. Fink, and A. Aubry, "Three-dimensional ultrasound matrix imaging," *Nat Commun*, vol. 14, no. 1, p. 6793, Oct. 2023, doi: 10.1038/s41467-023-42338-8.

[53] Z. Dong, U.-W. Lok, M. R. Lowerison, C. Huang, S. Chen, and P. Song, "Three-Dimensional Shear Wave Elastography Using Acoustic Radiation Force and a 2-D Row-Column Addressing (RCA) Array," *IEEE Trans. Ultrason., Ferroelect., Freq. Contr.*, vol. 71, no. 4, pp. 448–458, Apr. 2024, doi: 10.1109/TUFFC.2024.3366540.

[54] A. Salari, M. Audoin, B. Gueorguiev Tomov, B. Y. S. Yiu, E. Vilain Thomsen, and J. Arendt Jensen, "Beamformer for a Lensed Row–Column Array in 3-D Ultrasound Imaging," *IEEE Trans. Ultrason., Ferroelect., Freq. Contr.*, vol. 72, no. 2, pp. 238–250, Feb. 2025, doi: 10.1109/TUFFC.2025.3526523.